\documentclass[12pt]{article}

\usepackage{epsfig}
\usepackage{epstopdf}
\usepackage{fp}
\usepackage{color}
\usepackage{graphicx}
\usepackage[toc,page]{appendix}
\usepackage{natbib}
\usepackage{multicol}
\usepackage{amsmath}
\usepackage[margin=0.6in]{geometry}
\graphicspath{ {figs/} }

\newcommand{\Rs}{$ R_{\odot}$}

\newcommand{\de}{$^{\circ}$}

\begin{document}
\title{Automated detection of coronal mass ejections in three-dimensions using multi-viewpoint observations}
\author{Joe Hutton \& Huw Morgan}
\date{\small{Accepted for publication in Astronomy \& Astrophysics, 2016 November 24}}
\maketitle
\begin{center}
\small{Institute of Mathematics, Physics $\&$ Computer Sciences, Aberystwyth University, Penglais, Aberystwyth, Ceredigion, SY23 3BZ}\\ 
\small{joh9@aber.ac.uk}
\end{center}

\abstract
{A new, automated method of detecting coronal mass ejections (CMEs) in three dimensions for the LASCO C2 and STEREO COR2 coronagraphs is presented. By triangulating isolated CME signal from the three coronagraphs over a sliding window of five hours, the most likely region through which CMEs pass at 5\Rs\ is identified. The centre and size of the region gives the most likely direction of propagation and approximate angular extent. The Automated CME Triangulation (ACT) method is tested extensively using a series of synthetic CME images created using a wireframe flux rope density model, and on a sample of real coronagraph data; including halo CMEs. The accuracy of the angular difference ($\sigma$) between the detection and true input of the synthetic CMEs is $\sigma$=7.14$^{\circ}$, and remains acceptable for a broad range of CME positions relative to the observer, the relative separation of the three observers and even through the loss of one coronagraph. For real data, the method gives results that compare well with the distribution of low coronal sources and results from another instrument and technique made further from the Sun. The true three dimension (3D)-corrected kinematics and mass/density are discussed. The results of the new method will be incorporated into the CORIMP database in the near future, enabling improved space weather diagnostics and forecasting.\\\small{\textit{keywords} - Sun: corona---Sun: CMEs}}
\maketitle

\section{Introduction}
\label{sec:intro}
First observed by the \textit{Skylab} mission in the early 1970s, coronal mass ejections (CMEs) are the largest and most dynamic phenomena that originate from the Sun, and can be observed in the extended corona by white light coronagraphs (\citet{gosling74,webb12,chen11} and references within.). Huge eruptions of magnetised plasma, CMEs can propagate at speeds of up to thousands of kilometers per second and have a broad range of masses \citep{lui10,yashiro04}. Given that these eruptions and their associated bursts of energetic particles can have adverse effects such as geomagnetic storms at Earth, early warnings of their presence and their direction of propagation are needed for space-weather forecasting \citep{schwenn05}. Statistical information on CMEs is also invaluable for gaining a better understanding of their nature \citep{kwon14}.

For over a decade, CME events and charateristics (such as spatial size, velocity, acceleration, type, morphology and distribution) have been detected and catalogued using both manual and automated methods. The most widely used catalogues are: the Coordinated Data Analysis Workshop (CDAW\textsuperscript{1}) CME catalogue, the Computer Aided CME Tracking (CACTus\textsuperscript{2}) catalogue, the Solar Eruptive Event Detection System (SEEDS\textsuperscript{3}), Automatic Recognition of Transient Events and Marseille Inventory from Synoptic maps (ARTEMIS\textsuperscript{4}), and, more recently, the Coronal Image Processing (CORIMP\textsuperscript{5}) catalogue. These catalogues use data from the Large Angle Spectrometric Coronagraph (LASCO; \citet{brueckner95}) on board the Solar and Heliospheric Observatory (SOHO; \citet{domingo95}). The CDAW, CACTus and SEEDS catalogues all use running-difference techniques (where a previous image is subtracted from the current image) to locate CMEs in coronagraph images. This is a quick and easy way to effectively reveal the presence of a CME but is prone to several errors. Running differences do not show the true CME structure, but a time derivative. It is therefore difficult to use running-difference images for a structural interpretation of CMEs, as such spatio-temporal crosstalk can cause false artefacts to appear in the images. The CACTus and SEEDS catalogues employ intensity thresholding techniques (CACTus, in the Hough space) to detect and track the progression of the CME front, and hence speed, acceleration, angular span and core angles; whereas the CDAW Data Center CME catalgue is wholly manual in its operation, relying on user generated "point-$\&$-click" height-time plots for each event \citep{gopalswamy09, robbrecht04}. The ARTEMIS catalogues CME occurrences by their distinct signatures in Synoptic Maps of the corona for complete Carrington rotations, produced from the LASCO C2 coronagraph. The isolation at the CME signatures is based on adaptive filtering and segmentation, followed by merging with high-level knowledge. While the technique used accurately estimates parameters such as appearance time, position angle, angular extent, velocity, etc., the detections are severely biased by the viewing geometry and plane-of-sky (POS) approximations \citep{boursier09}. More recently, the CORIMP catalogue has been developed, which implements a dynamic background technique and multiscale edge detection \citep{morgan10,auto1}. These techniques isolate and characterise CME structure in the coronagraph field-of-view (FOV) and remove small-scale noise/features \citep{byrne09,auto2,byrne15}. However, all of these catalogues are limited to POS measurements. Regardless of the quality of the method, the POS approximation is always fraught with uncertainty and will always yield the lowest possible estimate for characteristics such as velocity and mass.

Since the advent of the Solar Terrestrial Relations Observatory (STEREO) mission in 2006 and the Sun-Earth Connection Coronal and Heliospheric Investigation (SECCHI) coronagraphs \citep{kaiser08,howard08}, numerous attempts have been made to track CMEs in three dimensions (3D). The twinned STEREO spacecraft (named Ahead (A) and Behind (B) ) provide unique perspectives of the Sun and extended corona and, for the first time, enables the study of CMEs in 3D \citep{mierla10}. Several previous studies have used techniques consisting of tie-pointing lines-of-sight across epipolar planes \citep{wood09,srivastava09,liewer09}. Tie-pointing techniques involve identifying the same feature (i.e. a specific plasma blob with a CME core) in separate, simultaneous observations and manually tracking just that small feature rather than the whole event. Methods of triangulating CME features using time-stacked intensity slices at a fixed latitude, colloquially named J-maps, have also been developed \citep{davis10}. J-map techniques have the same limitations as tie-pointing; namely that they do not consider the curvature of the feature, and the sight-lines may not intersect upon the surface of the observed feature. A technique for determining kinematic properties of CMEs has been established by \citet{davies13} which is based on a generalised self-similarly expanding circular geometry, named the Stereoscopic Self-Similar Expansion (SSSE) technique. SSSE also relies on the elongation profiles being extracted manually from J-maps using STEREO Heliospheric Imager 1 (HI-1) image data. The SSSE technique, as with all 3D analysis using HI-1/2, can only be carried out if the CME propagates along a trajectory between the two spacecraft, in the common FOV. The SSSE technique is the primary method used for the recent Heliospheric Cataloguing, Analysis and Techniques Service (HELCATS\textsuperscript{6}) CME kinematics catalogue. The 3D trigangulation technique of elliptical tie-pointing, developed by \citet{byrne10}, overcomes the limitations of previous tie-pointing techniques through fitting an ellipse into the quadrant where four sight-lines tangent to the leading edges of a CME intersect, as extracted manually in each image. Fitting ellipses in this way over many epipolar planes builds up a reconstruction of the CME front, and hence considers its curvature. Forward modelling of CMEs using a 3D flux rope based on a graduated cylinder model has also been applied, to some success, to STEREO observations \citep{thernisien09}. However, some of the parameters governing the model shape and orientation must be changed manually to best fit the twin observations simultaneously, which can often be a tedious process. The model is limited to the subset of large, well structured CMEs, and cannot be suitable for other CME structures \citep{hutton15}.

This study presents a new method for the simple, automated detection of a CME's longitude and latitude at 5\Rs\ using triangulated coronagraph observations from SOHO LASCO C2 and STEREO SECCHI COR2 named the Automated CME Triangulation (ACT) method. The ACT method is rigorously tested using a synthetic flux-rope CME model and comparisons with the low-coronal sources (flare, filaments etc.) of the CME, and the results presented in section \ref{sec:results}. Section \ref{sec:mass} discusses implications for CME mass and kinematics, and a summary is given in section \ref{sec:summary}.

\section{Method}
\label{sec:method}

\subsection{Identification of CME signal}
\label{sec:dst}
\begin{figure}[!t]
\centering
\includegraphics[width=6cm]{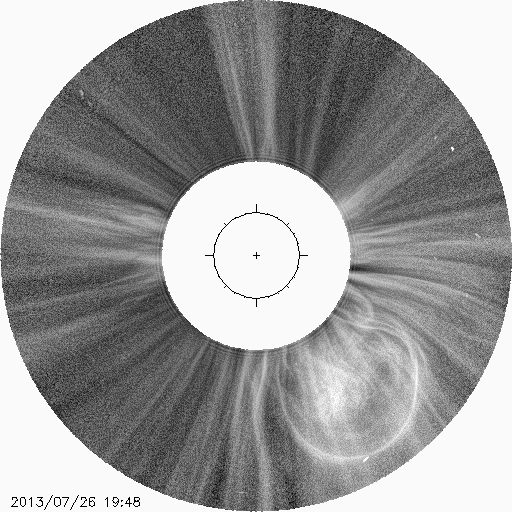}
\includegraphics[width=6cm]{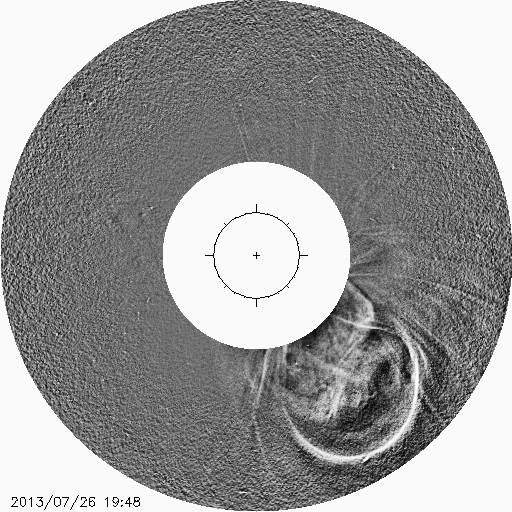}
\caption{CME observed by LASCO C2 on 2013-07-26 19:48. a) C2 image prior to DST, b) Separated dynamic component showing the main bright CME, other fainter dynamic events and noise.}
\label{fig:1}
\end{figure}

To detect the position of CMEs in 3D, observations from the SOHO LASCO C2 coronagraph are triangulated with those from the COR2 twin coronagraph instruments of the STEREO SECCHI spacecraft. LASCO C2 measures the Thomson-scattered emission from coronal electrons, as well as the unwanted signal from the dust F-corona \citep{morgan2007} and instrumental stray light. The spatial resolution of C2 is 11.4", with a useful FOV of 2.2 - 6.0\Rs. STEREO SECCHI COR2 coronagraphs provide a useful FOV of $\sim$3 - 14\Rs\  \citep{howard08}. In coronagraph images, CMEs are not viewed in isolation, but in the presence of the fine structural detail of the quiescent corona such as streamers or coronal holes. In order to accurately detect the pixels containing CME information in the coronagraph images, we use a processing technique, also used by the CORIMP CME catalogue, that is able to separate the dynamic CME signal from the background quiescent coronal structures, based on the assumption that the background coronal structures (streamers, coronal holes, etc.) are predominantly radial \citep{morgan06,morgan10,auto1}. The dynamic separation technique (DST) is based on spatial and time deconvolution. When applied to coronagraph observations, the clear structures of CMEs are revealed despite the presence of background structure that may be several times brighter than the CME. Figure \ref{fig:1} shows an example of the DST applied to a CME observed by LASCO C2 on the 26th July, 2013, from \citet{hutton15}. Figure \ref{fig:1}a shows the image prior to separation, and figure \ref{fig:1}b shows the dynamic component of the image post-separation.

\begin{figure}[!t]
\centering
\includegraphics[width=10cm]{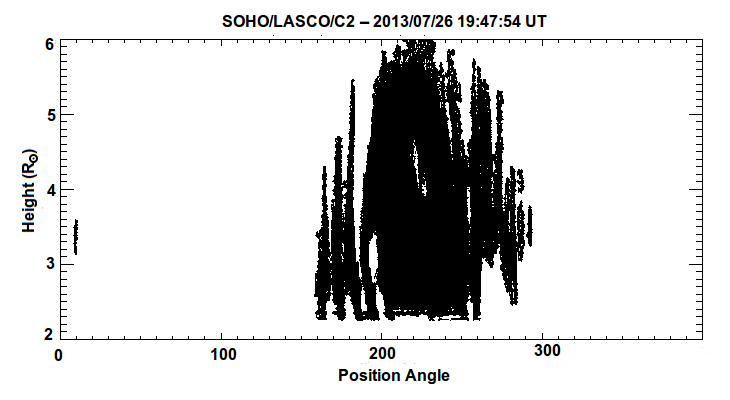}\\
\includegraphics[width=10cm]{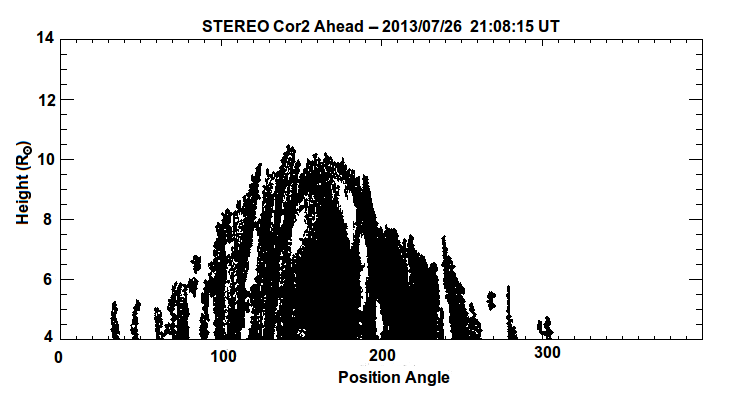}\\
\includegraphics[width=10cm]{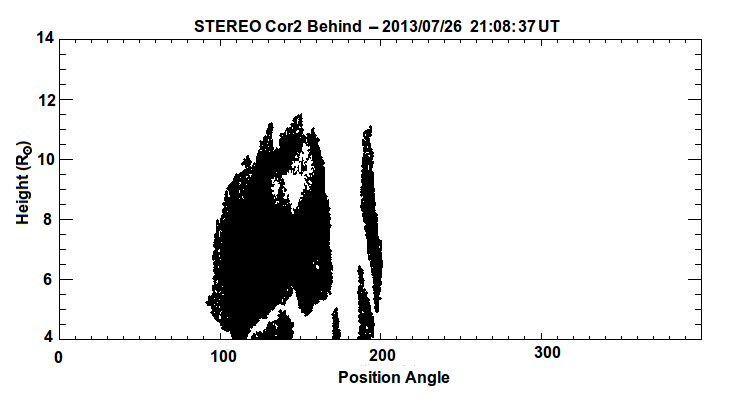}
\caption{Polar projection of the detected pixels containing CME information, from a CME occurring on the 26th July, 2013, as observed by a) SOHO/LASCO C2, b) STEREO Cor2 Ahead and c) Cor2 Behind.}
\label{fig:3}
\end{figure}

Figure \ref{fig:3}a-c shows examples of DST-processed images in polar coordinates (in the POS for each instrument, where a position angle of 0\de\ represents solar north) for a CME on the 26th July, 2013, as observed by LASCO C2 and STEREO Cor2 Ahead/Behind coronagraphs. CME signal in DST images is significantly above the level of the background noise, and this allows a straightforward detection of points which contain CME signal. Without the removal of background structure, this would be impossible. CME points are identified by:
\begin{itemize}
\item `Flattening' the image to reduce the intrinsic decrease of signal with coronal height. This is done by dividing by the mean signal at each height bin, calculated across all position angles.
\item Create a binary mask image where pixels with value greater than a certain threshold are set to one, and the remaining pixels are set to zero.
\item Identify connected groups of detection pixels. Groups with a number of pixels (or detector area) less than a certain threshold are removed from the binary mask. This is a step that very effectively removes noise from the detections.
\item Record detected pixels within the mask. Other pixels are discarded.
\end{itemize}
The set of detected pixels are used for the backprojection process. Detected points are recorded from images collected by multiple coronagraphs over a 5-hour time period. To find the region through which the CME passed at a specified distance above the solar surface, only CME points detected within the 4.5-5.5\Rs\ height range are kept for further analysis, as these heights are within the common FOV of the three coronagraphs. 


\subsection{Synthetic flux-rope CME}
\label{sec:synthcme}
\begin{figure}[!t]
\centering
\includegraphics[width=4cm]{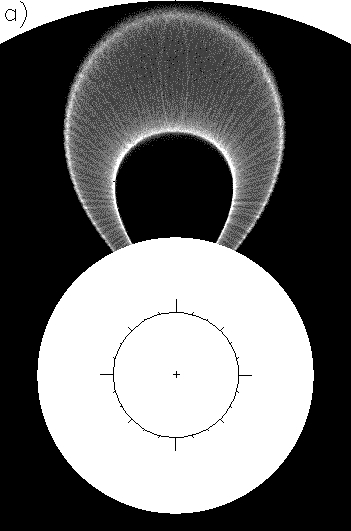}
\includegraphics[width=4cm]{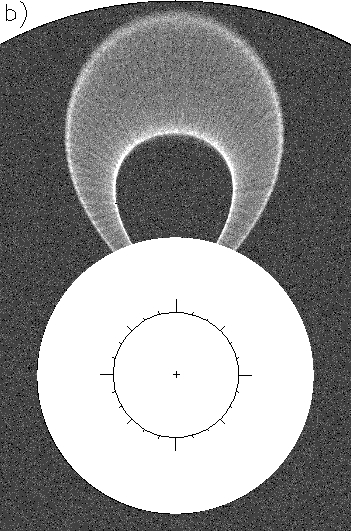}
\caption{Example synthetic flux-rope model. a) A simple rendering of the wire frame structure where a volume within a small distance from the surface of the CME model contains a random distribution of points to imitate Thompson scattering. b) The same synthetic CME with random noise added to the background. Examples are rendered as if observed by LASCO C2. Inner and outer circles represent the sizes and positions of the Sun and the C2 occulting disk respectively.}
\label{fig:2}
\end{figure}

A density model is used to create synthetic data for testing of the CME detection method. The model is very successful at replicating the appearance of flux-rope CMEs in coronagraph images (see e.g. \citet{auto1}). Similar to \citet{thernisien09}, the model consists of a long tubular shape created from a continuous helix of varying radius and  specified number of rotations, with two foot points anchored to the photosphere that create the body of the CME. To imitate Thompson scattering, a volume within a small distance from the surface of the model now contains a random distribution of points. These points are then treated as individual scattered electrons (Figure \ref{fig:2}a). For a fuller description of the model see \citet{hutton15}. The wire-frame CME shown in figure \ref{fig:2} has been rendered as if observed by SOHO/LASCO C2, with the sizes and locations of the Sun and coronagraph occulter represented by the inner and outer circles respectively. Noise has been added to the background of the model in figure \ref{fig:2}b. Adding noise means that an automated CME detection procedure would have to identify the synthetic CME above this background, just as it would have to for actual coronagraph images after having been processed using DST.

\subsection{The Automated CME Triangulation (ACT) method} 
\label{sec:detect}

Each image pixel containing a detected CME signal corresponds to an extended line of sight (LOS). The true distribution of the CME material along the LOS is not known. Geometrically, the pixel position is transformed into a large set of points, back-projected along the LOS into the reconstruction space. Image pixels are converted to spherical Carrington coordinates by first defining a spherical coordinate system with the origin at the observing spacecraft (i.e. SOHO or STEREO A/B), with the z-axis directed towards the solar centre. Points are distributed along the LOS up to $\pm$5\Rs\ from the point of closest approach to the Sun. The positions of these points are then converted into Heliographic Radial-Tangential-Normal (HGRTN) coordinates (coordinate system centred on the Sun with the x-axis pointing through the observing spacecraft \citep{thompson06}), and from HGRTN converted into spherical Carrington coordinates. This process is repeated for each pixel in the DST processed  images taken over 5 hours from the three coronagraphs (LASCO C2 and SECCHI COR2 A and B) leading to a cloud of points in Carrington coordinates. Each point is a potential candidate for containing CME signal. The method works by ignoring the time coordinate, taking a 5-hour set of detections, and also ignoring the height (radial) component. The points are thus binned into a regular longitude-latitude grid. 

The final step is to identify the grid bins found to contain CME detections by all three coronagraphs. For each viewpoint, each point along the LOS is assigned its original DST-processed intensity, and the points found not to contain CME material are assigned a zero value. The intensities from each observer are multiplied together for each coordinate in the new system. Only the region identified as containing CME material by all three observers remains non-zero. This is taken to be the most probable region through which the CME passes from 4.5-5.5 \Rs. The multiplication of intensities gives a ``score'' , which can be mapped to the region of interest (ROI), shown in grey-scale in later figures, indicating areas within the ROI through which more CME material is likely to have passed. The result of this method is a longitude-latitude mapping of scores that indicate the most likely passage of a CME. The centre of the ROI is assumed as the centroid of the CME, and therefore the line along which the CME propagates in three dimensions. An estimate of median longitude-latitude is improved through a simple iterative process, leading to a robust estimate of the CME centroid.

\section{Results}
\label{sec:results}
\subsection{Testing using synthetic CME model}
\label{sec:testing}
\begin{figure}[!t]
\centering
\includegraphics[width=6cm]{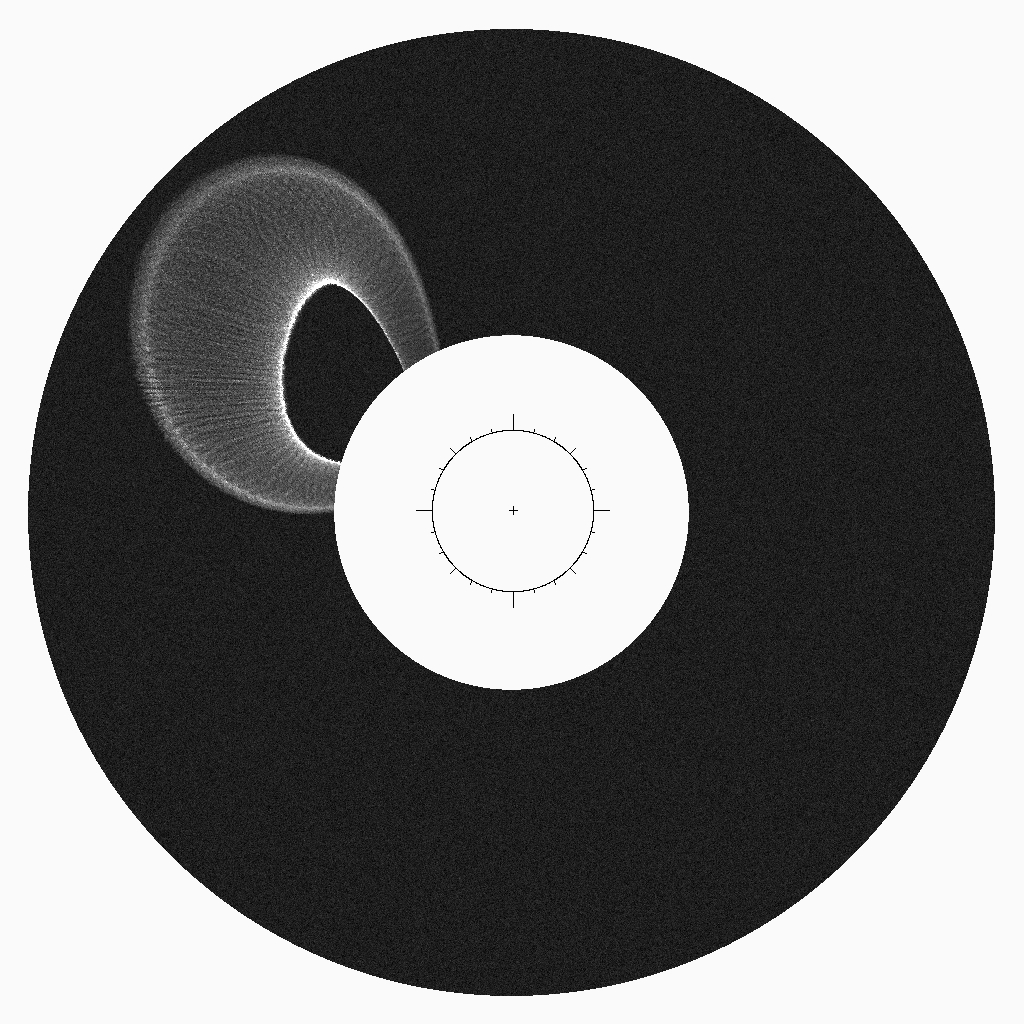}
\includegraphics[width=6cm]{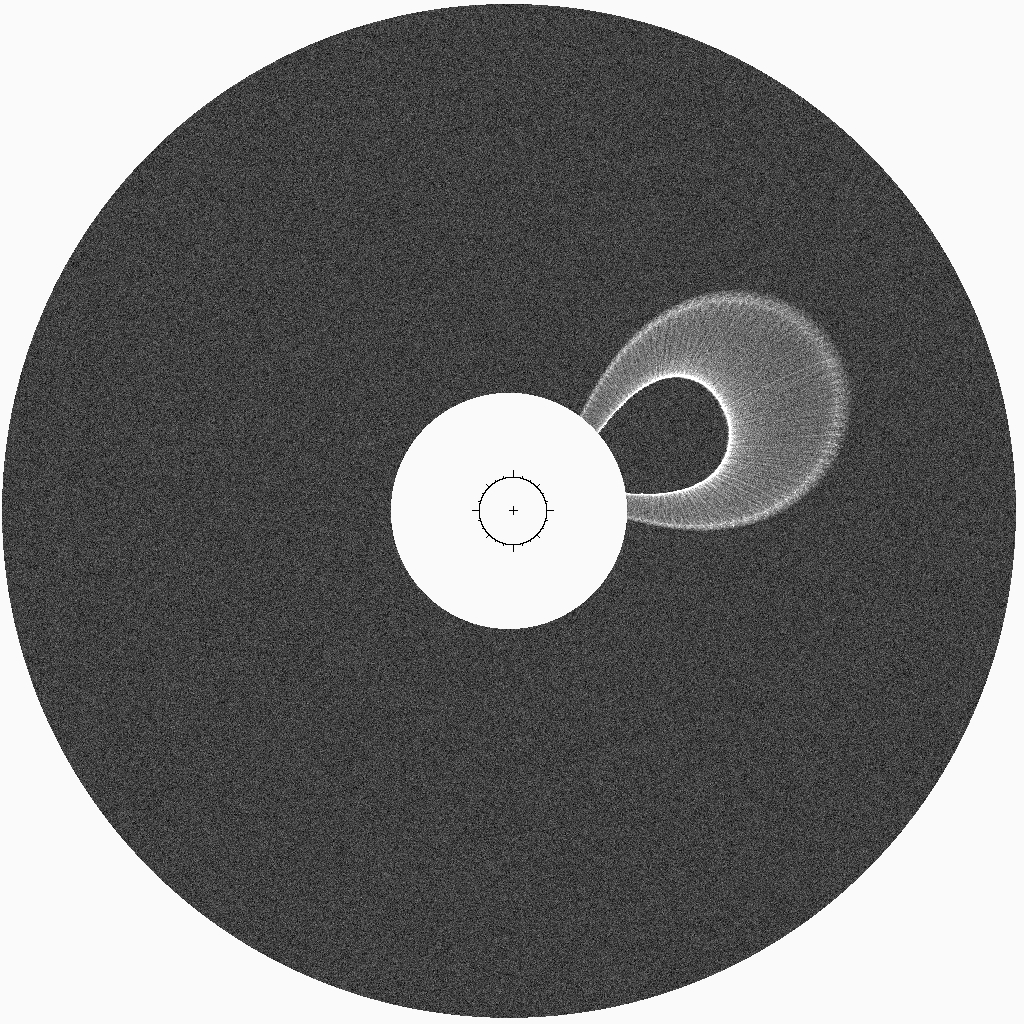}\\
\includegraphics[width=6cm]{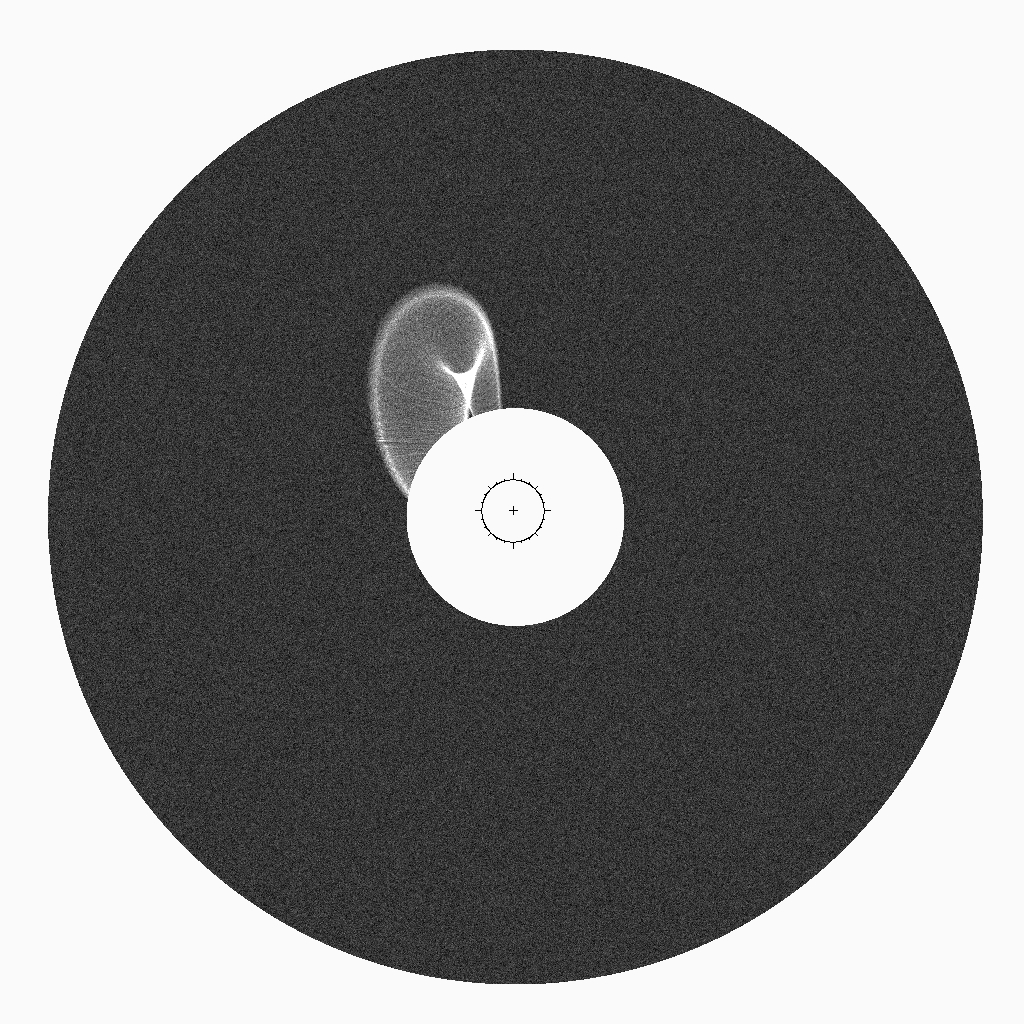}
\includegraphics[width=6cm]{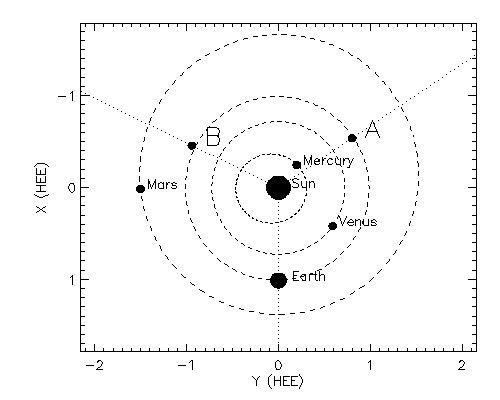}
\caption{Example of a synthetic CME rendered from a flux rope density model. Modelled to have occurred on the 1st September, 2012 at T02:36. Rendered from the perspective of the three observing spacecraft/instruments: a) STEREO COR2 Ahead, b) STEREO COR2 Behind, and c)SOHO/LASCO C2. d) Orbital plot showing the relative positions of the three spacecraft.}
\label{fig:4}
\end{figure}

\begin{figure}[!t]
\centering
\includegraphics[width=8.5cm]{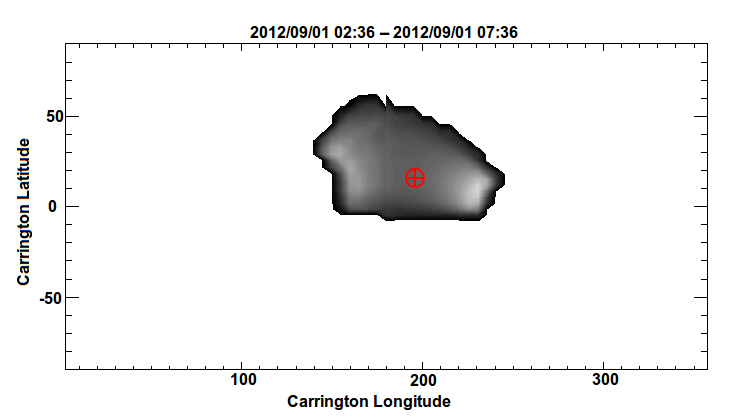}\\
\includegraphics[width=8.5cm]{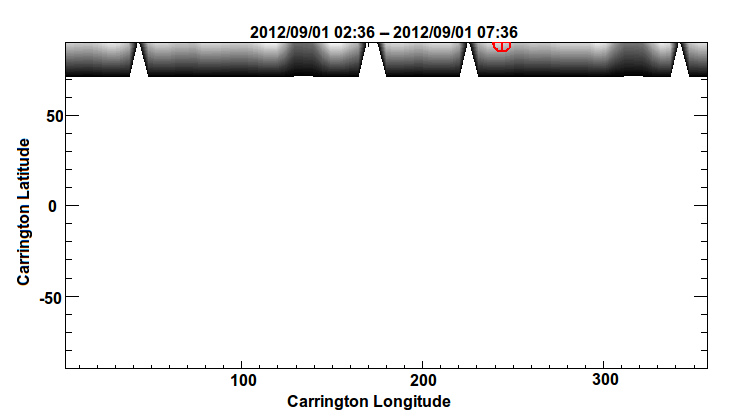}\\
\includegraphics[width=8.5cm]{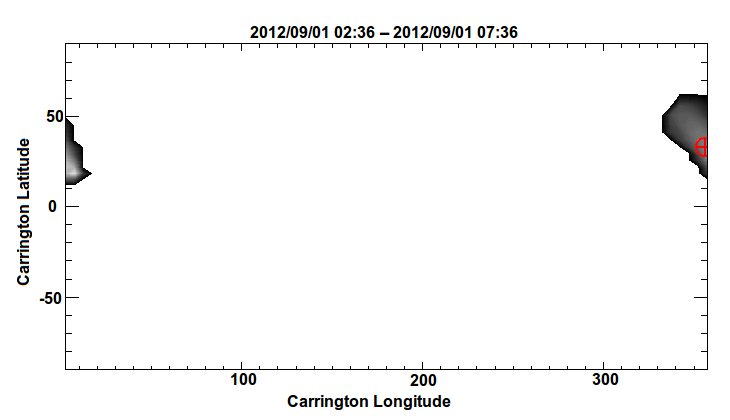}
\caption{Maps of the detected regions through which a CME passed in Carrington coordinates. a) Example event corresponding to Figure \ref{fig:4}. b) CME erupting at a high latitude. c) CME that erupted from close to 0$^{\circ}$ Carrington longitude. Cross-hairs represent the detected centroid.}
\label{fig:5}
\end{figure}

To test the accuracy of the ACT method, a series of synthetic CMEs, as described in section \ref{sec:synthcme}, is input to the procedure. Figure \ref{fig:4} is an example of a synthetic CME modelled for 2012-09-01T02:36, as seen by a) STEREO Cor2 Ahead, b) STEREO Cor2 Behind and c) SOHO C2. Figure \ref{fig:4}d shows the relative positions of the three observers at the modelled date/time. This date was chosen as the spacecrafts were separated from one another by close to 120$^{\circ}$ ( $\widehat{LA}$:124$^{\circ}$, $\widehat{LB}$:116$^{\circ}$, $\widehat{AB}$:120$^{\circ}$, where $L$, $A$ and $B$ denote the positions of LASCO, STEREO Ahead and Behind respectively.) giving an ideal configuration. Figure \ref{fig:5}a shows the results of the detection, displayed in a Carrington longitude vs. latitude map. The centre of the ROI, and hence the assume centroid of the CME's propagation, is represented by the cross-hairs. The grey-scale represents the intensity score described in section \ref{sec:detect}. The accuracy of the detection is tested by comparing the longitude/latitude results to those used as input parameters for the synthetic CME. These results are shown in Table 1. Two additional cases are tested: A CME erupting at a high latitude close to the solar north pole, and a CME erupting close to 0$^{\circ}$ Carrington longitude, such that the angular extent of the CME crosses either side of the detection map. These configurations could potentially pose problems. Both of these cases are also modelled for an observation on the 1st September, 2012, and results of these detections are shown in figure \ref{fig:5}b \& c and Table 1. The accuracy is determined through examination of the angular difference between the input parameters and returned centroid. This angle is calculated by: 

\begin{equation}
\sigma = \arccos\left[ \sin(\phi_{1})\sin(\phi_{2}) + \cos(\phi_{1})\cos(\phi_{2})\cos(\theta_{1}-\theta_{2}) \right],
\label{eq:3}
\end{equation}
where $\phi_{1,2}$ and $\theta_{1,2}$ are the latitudes and longitudes of the two coordinates, and \texttt{$\sigma$} is the angular difference. The angular difference for the three example cases ranges from 6.0$^{\circ}$ $\le$ $\sigma$ $\le$20$^{\circ}$.

\begin{table}
\label{table:1}
\centering
\caption{Table of results showing the comparisons between the input longitude and latitude ($\theta_{in}$, $\phi_{in}$) of the synthetic CME model, the detected CME centres ($\theta_{out}$, $\phi_{out}$) and the angular distance ($\sigma$) between these two coordinates. All results are shown in degrees Carrington.}
\begin{tabular}{lrrrrr}
 & $\theta_{in}$ & $\theta_{out}$ & $\phi_{in}$ & $\phi_{out}$ & $\sigma$ \\\hline
Initial Example: & 169 & 170 & 18 & 38 & 20.0\\
High Latitude: & 148 & 244 & 84 & 90 & 6.0\\
Carrington Zero: & 0 & 357 & 18 & 31 & 13.3\\
\end{tabular}
\end{table}

\subsection{Spacecraft separation}
\label{sec:separation}
The separation angle of the three observers is constantly changing as STEREO A/B orbit in opposite directions (relative to Earth) around the Sun at a rate of $\pm$22$^{\circ}$ per year \citep{kaiser08}, while SOHO remains at the first Lagrangian (L1) point \citep{domingo95}. The ACT detections will be presumably most accurate whilst the three observers form an approximate equilateral triangle around the Sun ( i.e. separated from the other spacecraft by longitudes of 60 or 120$^{\circ}$) as was the case in September, 2012. Another arrangement that lends itself to accurate 3D event detection is while the observatories are in quadrature ( i.e. when the STEREO crafts are at 45$^{\circ}$ angular separation from SOHO and the Sun–Earth vector, and at 90$^{\circ}$ from each other) as in December, 2008. During the period approaching March, 2015, the STEREO spacecraft were in conjunction at the far side of the Sun, that is, directly opposite SOHO along the Sun-Earth vector. It is anticipated that any automated 3D detections during this period would have suffered from reduced accuracy as the STEREO spacecraft would have been observing from the same angle, effectively reducing the number of observers to two. The effect of the observing angles on the accuracy of ACT is tested on synthetic CME data in section \ref{sec:testing2}. Unfortunately, on the 1st October, 2014, all contact with the STEREO-B spacecraft was lost, reducing the number of viewpoints to two.  On the 21st of August, 2016 contact was briefly re-established with STEREO-B, and an unsuccessful recovery attempt was made, however at the time of publication, no observational data has been received from the spacecraft since the initial contact loss. The loss of STEREO-B is discussed in section \ref{sec:loss}.

\subsubsection{Testing on synthetic data}
\label{sec:testing2}
\begin{figure}[!t]
\centering
\includegraphics[width=8.5cm]{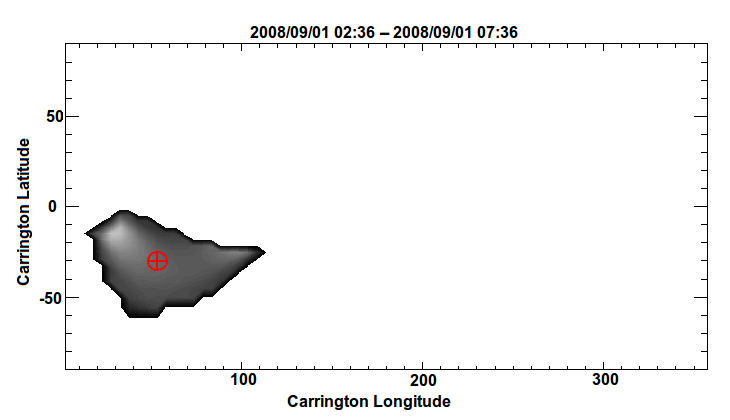}
\includegraphics[width=8.5cm]{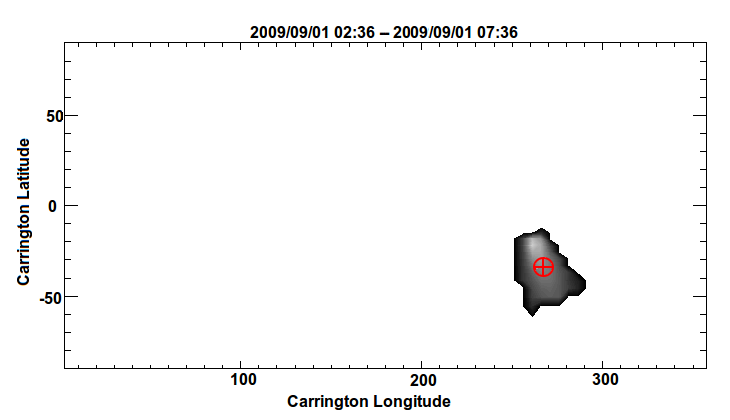}\\
\includegraphics[width=8.5cm]{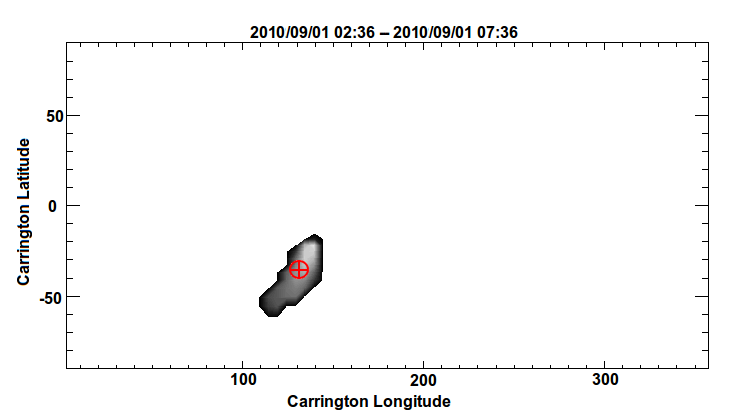}
\includegraphics[width=8.5cm]{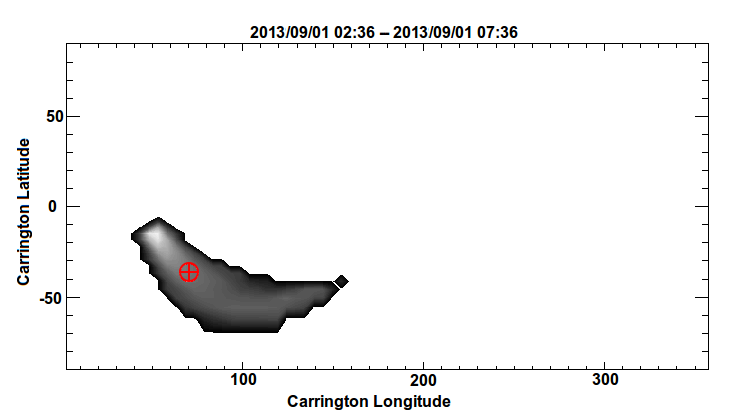}
\caption{Maps in Carrington coordinates of the detected regions through which identical CMEs passed, but having been modelled for the changes in observer separation from the years 2008, 2009, 2010 and 2013 (a-d, respectively). Cross-hairs represent the location of the centroid.}
\label{fig:6}
\end{figure}

To test the method against spacecraft angular separation, a series of synthetic CMEs were created with identical size, velocity and latitudes and longitudes of origin relative to Earth. However, all CMEs were modeled such that they occurred on different dates, ranging from 2008 to 2013. Again, by comparing the input latitude and longitude parameters of the synthetic CME to the output coordinates of the ACT detections, the effect of spacecraft separation on the accuracy of the detection can be determined. Each synthetic CME is rendered for a 1$^{st}$  Sept observation. Figure \ref{fig:6}a-d shows the results of the detections, and Table 2 contains the comparison with the synthetic CME model input parameters. The angular difference between input and output conform with the range found for the example cases in section \ref{sec:testing}; $\sigma$ = 2.7$^{\circ}$ and 4.9$^{\circ}$ in 2013 and 2009, respectively, and the average, $\bar{\sigma}$=3.9$^{\circ}$. The chosen input parameters for the wireframe model meant the synthetic CME was rendered to be on the far side of the Sun during the 2011-2012 observations, therefore could not be seen in the STEREO-B COR2 FOV. The results for this scenario will be examined in section \ref{sec:loss} where the effects of the loss of an observer are examined in more detail.

\begin{table}
\label{table:2}
\centering
\caption{As Table 1, listing the affect on the accuracy of the detection method due to the change in spacecraft separation through the years 2008 to 2013. All results are shown in degrees Carrington.}
\begin{tabular}{lrrrrr}
 & $\theta_{in}$ & $\theta_{out}$ & $\phi_{in}$ & $\phi_{out}$ & $\sigma$ \\\hline
2008: & 50 & 54 & -34 & -31 & 4.9\\
2009: & 273 & 267 & -34 & -34 & 4.7\\
2010: & 135 & 131 & -34 & -35 & 3.4\\
2013: & 69 & 70 & -34 & -37 & 2.7\\
\end{tabular}
\end{table}

\subsubsection{Loss of an observatory}
\label{sec:loss}
There are several scenarios in which a CME might not be viewed by all three observers: A CME could erupt on the far side of the Sun from one of the observatories, causing the CME to be obscured by the coronagraph's occulting disk; the CME angle of eruption relative to the observers could cause the CME not to pass through the defined POS height range during the same time interval for each data; or there could be a gap in the data set for one or more observer. In this section, synthetic CMEs are detected using data from only two coronagraphs to test the influence of missing data from an observatory data set. With the present lack of observational data being received from STEREO-B, testing in this way is even more important. If further attempts to recover STEREO-B also fail, all future use of this technique hinges on the ability to function with only two coronagraphs. The method is tested on the same 2008-2013 synthetic examples as section \ref{sec:testing2} but using only the data from LASCO C2 and STEREO-A. Table 3 contains the comparison with the synthetic CME model input parameters. The average angular difference between input and output is $\bar{\sigma}$=11.5$^{\circ}$, significantly larger than $\bar{\sigma}$=3.9$^{\circ}$ found in section \ref{sec:testing} ($\Delta\bar{\sigma}$=7.6$^{\circ}$). Results range from 6.1$^{\circ}$ for the 2009 simulation, to 16.9$^{\circ}$ in 2013. This pattern aligns with configurations for favourable triangulation; with the most accurate results being returned when the observers are in 2009, and the least accurate when STEREO Ahead the opposite side of the Sun from LASCO in 2013.

\begin{table}
\label{table:3}
\centering
\caption{As Table 2, with the loss of one of the three observing coronagraphs, STEREO Behind. All results are shown in degrees Carrington.}
\begin{tabular}{lrrrrr}
 & $\theta_{in}$ & $\theta_{out}$ & $\phi_{in}$ & $\phi_{out}$ & $\sigma$ \\\hline
2008: & 50 & 60 & -34 & -30 & 9.9\\
2009: & 273 & 267 & -34 & -31 & 6.1\\
2010: & 135 & 149 & -34 & -34 & 11.2\\
2011: & 357 & 345 & -34 & -30 & 11.3\\
2012: & 207 & 196 & -34 & -24 & 13.6\\
2013: & 69 & 53 & -34 & -25 & 16.9\\
\end{tabular}
\end{table}

\subsection{Testing on real CME data}
\label{examples}
In this section the ACT method is tested on real coronagraph data from LASCO C2 and STEREO COR2. Three example CME events are selected from the CORIMP CME catalogue: \emph{Event 1}: CME erupting from the east limb as seen by C2, close to the equator, on the 14th January, 2009. The time limits for the detection covers the 5 hour sliding window from 07:00 to 12:00 UT. This CME appears quite small in it's spatial extent, with no clear structure. \emph{Event 2}: 26th October, 2013, from 13:25 to 18:25 UT. CME has a clear three-part structure: A bright leading front, followed by a cavity and a bright central core \citep{chen11,vourlidas2013}. \emph{Event 3}: 27th January, 2012 from 18:30 to 23:30 UT. CME appeared as a partial halo in C2 data. The centroid position results of the ACT method are compared to those in the HELCATS WP3 catalogue and/or the location of the CME triggering event (e.g. filament eruption or flare) from the Heliophysics Events Knowledgebase (HEK\textsuperscript{7}) \citep{hulburt12} where available. The triggering events from HEK are readily matched with the CMEs manually through visual comparisons and temporal correlations. All coronagraph data were processed with DST to remove the quiescent background (as described in section \ref{sec:dst}).

\subsubsection{Event 1 CME: 14th January, 2009}
\label{sec:smaller}
\begin{figure}[!t]
\centering
\includegraphics[width=6cm]{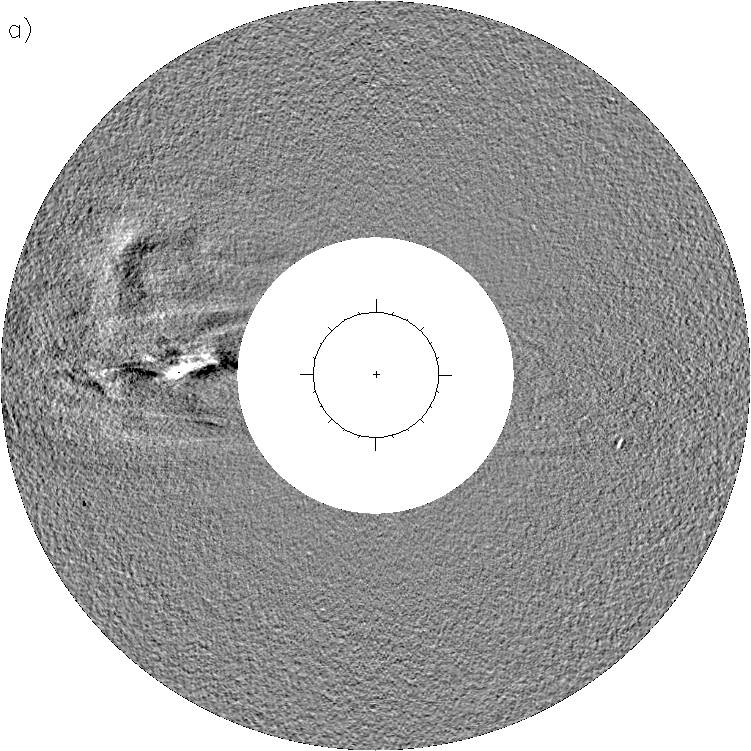}
\includegraphics[width=6cm]{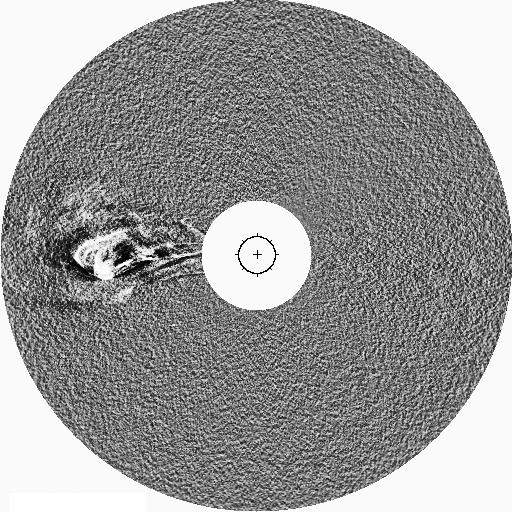}\\
\includegraphics[width=8.5cm]{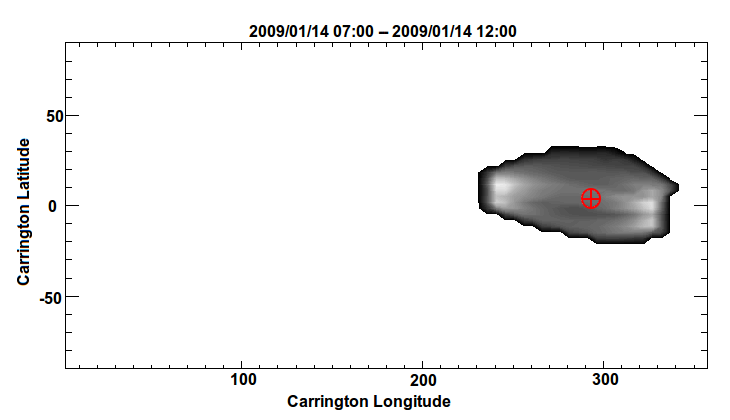}\\
\caption{Event 1: CME observerd on the 14th January, 2009, from 07:00 to 12:00. a) LASCO C2, b) STEREO Cor2 Ahead, images processed using DST. c) Detected most probable region through which the CME passed at 5\Rs. Cross-hairs indicate the centroid.}
\label{fig:7}
\end{figure}

The CME on the 14th Janurary, 2009 appears quite small in terms of spatial extent, as seen by LASCO C2 (figure \ref{fig:7}a, processed with DST) and STEREO Cor2 Ahead (figure \ref{fig:7}b). Figure \ref{fig:7}c shows the results of the detection procedure as a Carrington latitude/longitude map. The result shows a large ROI, which corresponds to the east solar limb from the C2 POV, and indicates the most probable region through which the CME propagated. The centroid of the CME is calculated to be (293$^{\circ}$, 4$^{\circ}$). The result for the same CME as listed in the HELCATS WP3 catalogue, from STEREO HI-1
data, is (276$^{\circ}$, 12$^{\circ}$), which gives an angular separation of $\sigma$=18.6$^{\circ}$. This difference is larger than those produced through the wire-frame CME model images (sections \ref{sec:testing}-\ref{sec:separation}), which is to be expected when making the leap from a synthetic model to real data. However, the difference may simply be due to systematic differences between the methods.

\subsubsection{Event 2, 3-part CME: 26th October, 2013}
\label{sec:3part}
\begin{figure}[!t]
\centering
\includegraphics[width=6cm]{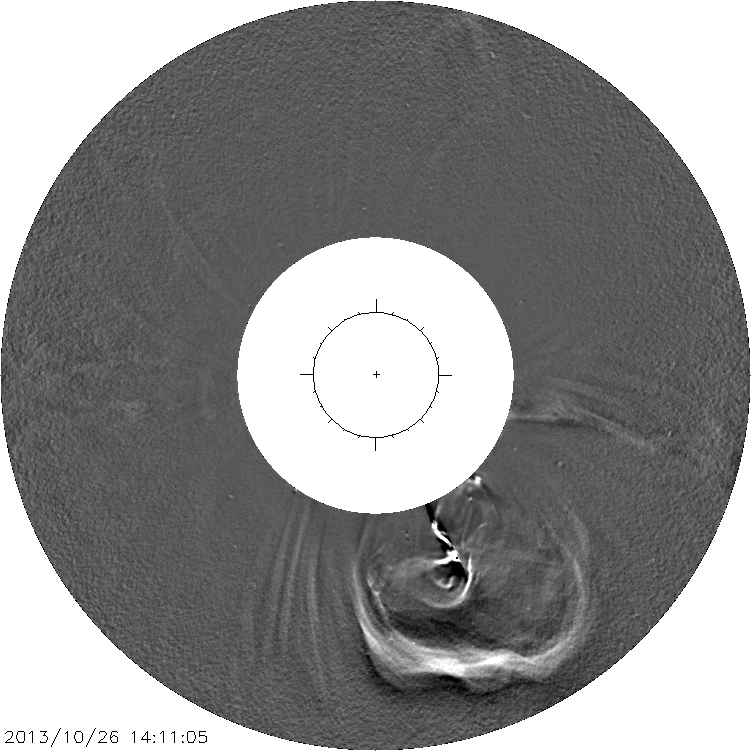}
\includegraphics[width=6cm]{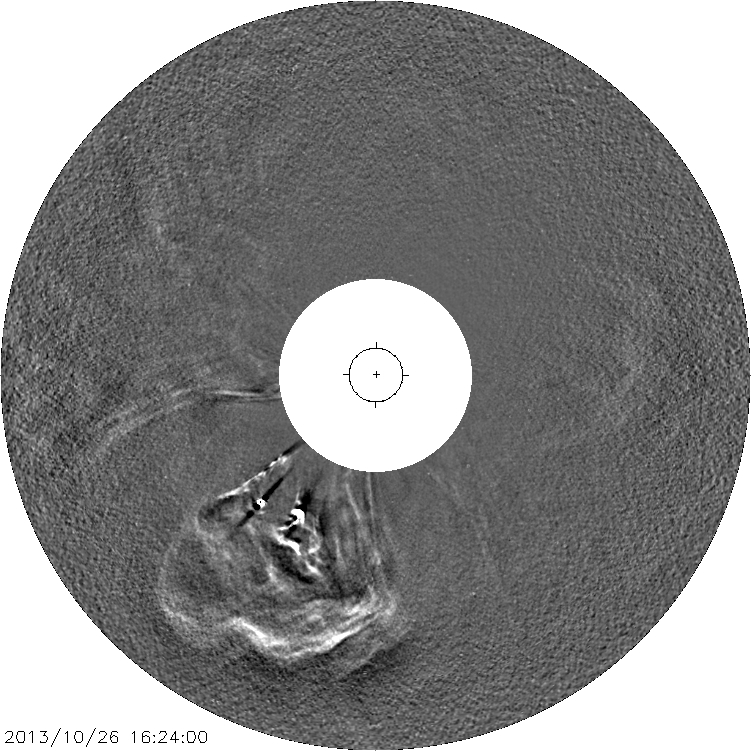}
\includegraphics[width=6cm]{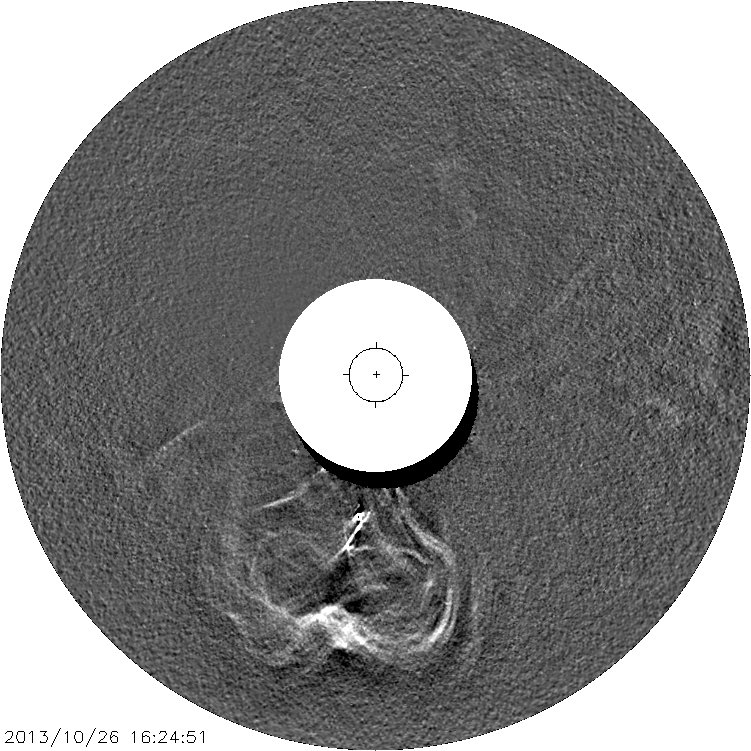}\\
\includegraphics[width=8.5cm]{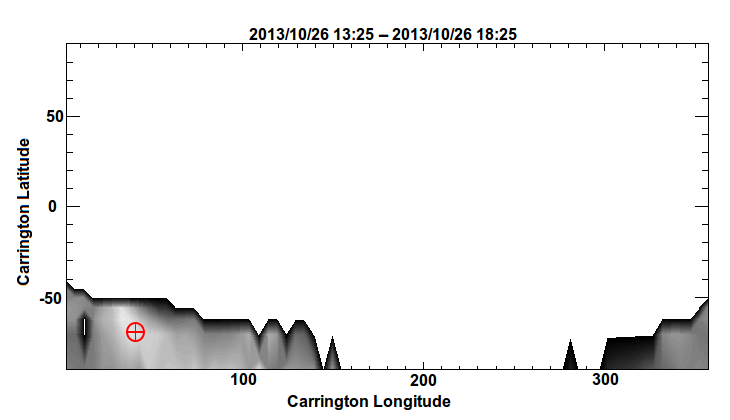}
\caption{Event 2: CME observerd on the 26th October, 2013, from 13:20 to 18:20. a) LASCO/C2 (26/10/2013 14:11), b) STEREO A COR2 (16:24), c) STEREO B COR2 (16:24) images processed using DST. d) Detected most probable region through which the CME passed at 5\Rs. Cross-hairs indicate the centroid.}
\label{fig:8}
\end{figure}

\begin{figure}[!t]
\centering
\includegraphics[width=12cm]{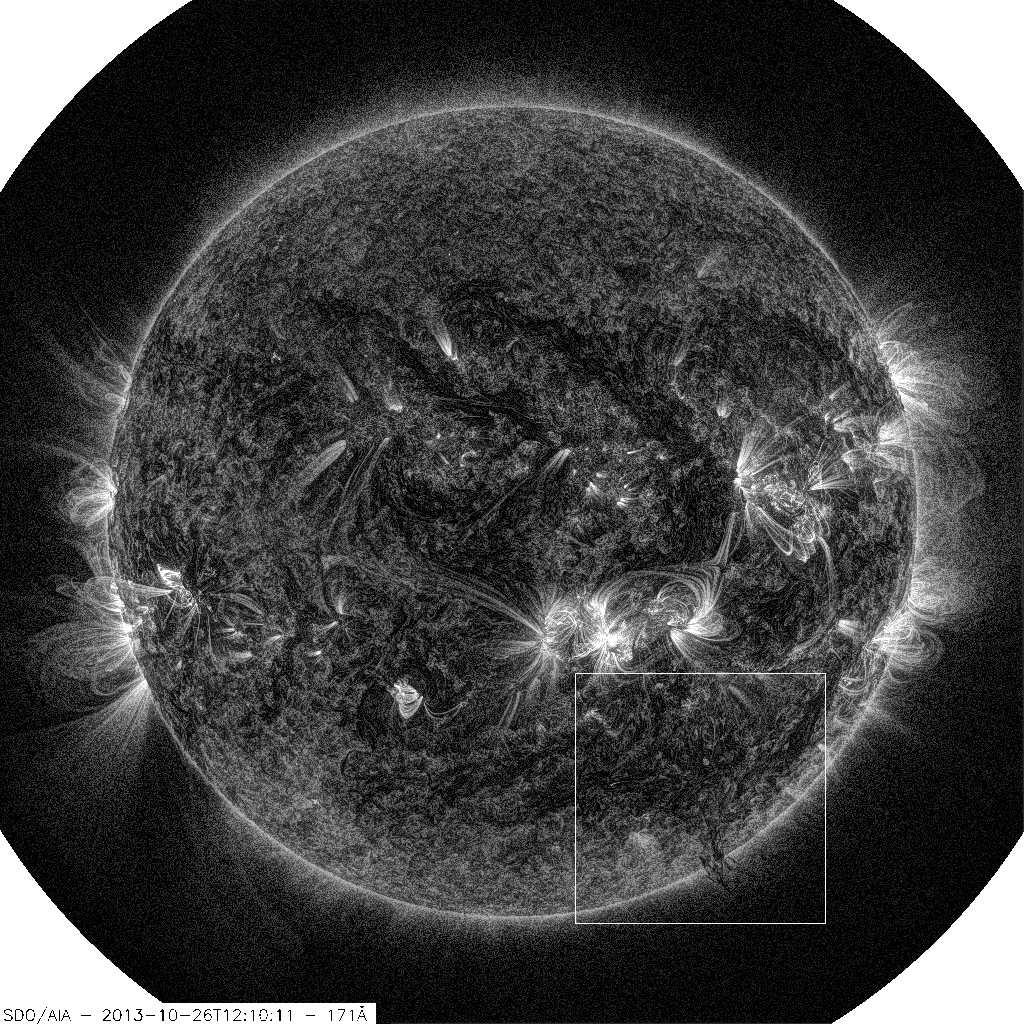}
\caption{SDO/AIA 17.1nm bandpass image of the filament eruption which triggered the Event 2 CME. Image date/time: 26/10/2013 12:10 UT. Processed using Multiscale Gaussian Normalisation.}
\label{fig:9}
\end{figure}

\begin{figure}[!t]
\centering
\includegraphics[width=12cm]{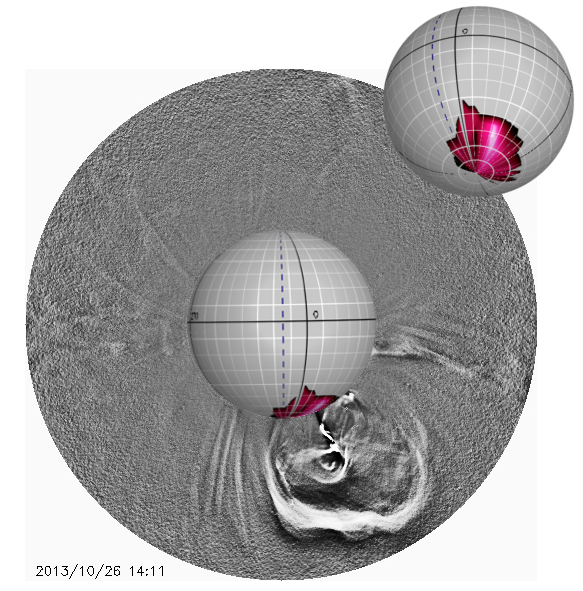}
\caption{Results of the detection for the CME on the 26th October,  2013 (Figure \ref{fig:8}d) plotted onto the surface of a sphere and set into the centre of Figure \ref{fig:8}a to give context to the result. The meridional longitude is shown by the dashed line. The same spherical plot, orientated to better show the region of interest, is set into the top corner of the image.}
\label{fig:10}
\end{figure}

One of several CMEs to occur on this date, this CME has a clear three-part structure, and erupted near the south pole as seen by all three coronagraphs. Figure \ref{fig:8}a-c shows the CME as seen by LASCO/C2, and STEREO COR2 A/B, respectively, processed with DST. Figure \ref{fig:8}d shows the results of the detection. The centroid is calculated to be (52$^{\circ}$, -70$^{\circ}$). This CME is not listed on HELCATS WP3 as the CME did not appear in the STEREO HI FOV. Therefore the result shall be compared to the location of the filament eruption which triggered this CME. Figure \ref{fig:9} is an image of this filament created using data from the Atmospheric Imaging Assembly on board the Solar Dynamics Observatory (AIA/SDO, \citet{lemen11}) using the 17.1nm bandpass filter at 12:10 UT. This image is processed using the Multiscale Gaussian Normalisation method \citep{mgn}. The location of the filament's eruption from the solar surface is given as (37$^{\circ}$, -52$^{\circ}$) by HEK. Figure \ref{fig:10} shows the results of the detection (Figure \ref{fig:8}d) plotted onto the surface of a sphere and set into the centre of Figure \ref{fig:8}a to give context to the result. The Earth-Sun vector is shown by the dashed line. The same spherical plot, orientated to better show the region of interest, is set into the top corner of the image. This visual comparison demonstrates the strong visual correlation between the  returned ROI and the location of the CME as seen by the coronagraph. The angular difference between our calculation of the centroid and that of the filament from HEK is calculated as $\sigma$=19.6$^{\circ}$. The problem with comparing the trigger event in this way is that such a comparison assumes a CME propagates radially from this point. In fact, non-radially propagating CMEs are often observed, and are well documented (e.g. \citet{liewer15}). Such deflections are thought to occur due to local uneven magnetic pressure forces, such as the polar coronal holes, causing an asymmetric expansion.

\subsubsection{Event 3, Halo CME: 27th January, 2012}
\label{sec:halo}
\begin{figure}[!t]
\centering
\includegraphics[width=6cm]{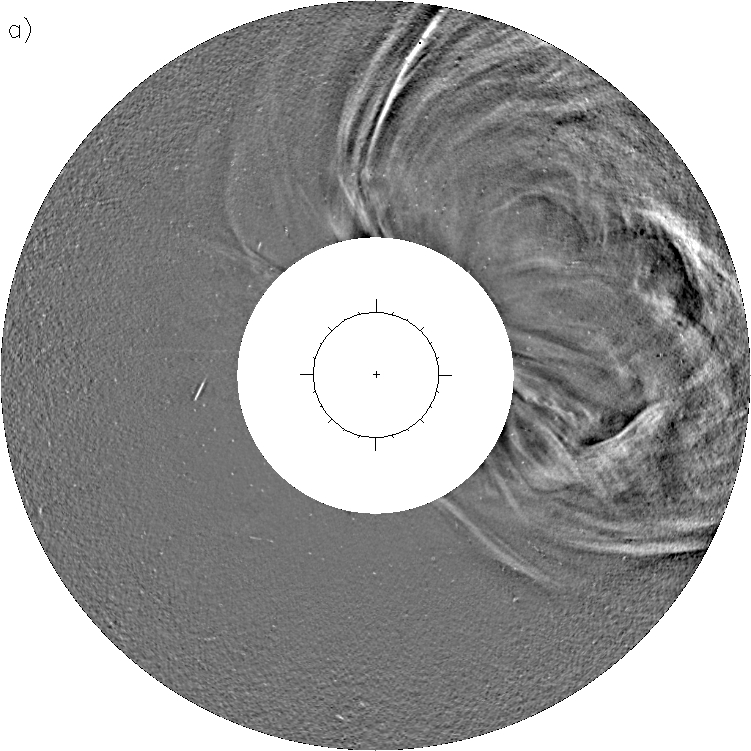}
\includegraphics[width=8.5cm]{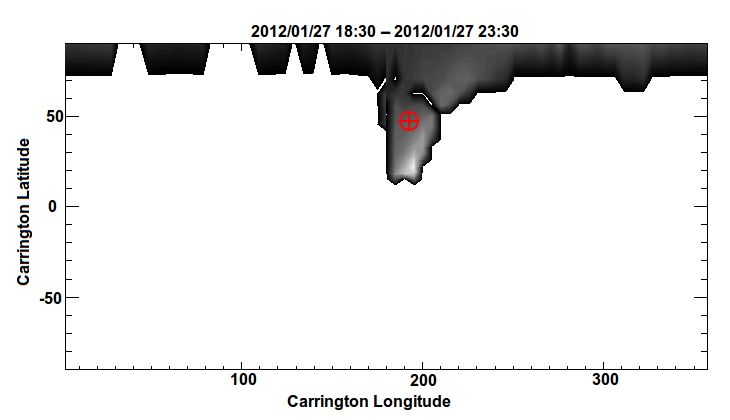}\\
\caption{Event 3: Partial Halo CME from the LASCO/C2 POV, observed on the 27th January, 2012, from 18:30 to 23:30. a) LASCO C2 image processed using DST. b) Detected most probable region through which the CME passed at 5\Rs. Cross-hairs indicate the CME's centroid.}
\label{fig:11}
\end{figure}

The CME occurring on the 27th January, 2012, appeared as a partial halo as seen by LASCO. Halo CMEs are a subset of CMEs that expand rapidly and appear to surround the occulting disk of the observing coronagraphs due to the propagation direction being along (or near to in the case of partials) the meridian between source and observer \citep{howard82}. Figure \ref{fig:11}a is a LASCO/C2 image of the CME at 19:00 UT. Figure \ref{fig:11}b is the result of the detection. Despite the CME occupying a greater portion on the C2 FOV (due to haloing), the centroid of the CME is reliably calculated to be (192$^{\circ}$, 47$^{\circ}$). The filament eruption that triggered this CME was identified by HEK to have erupted from (216$^{\circ}$, 30$^{\circ}$), which represents an angular difference from our result of $\sigma$=8.0$^{\circ}$. HELCATS estimate the CME to propagate along the (180$^{\circ}$, 31$^{\circ}$) line; representing an angular difference of $\sigma$=23.4$^{\circ}$. This difference is large, and reflects the inherent uncertainties involved in each of the methods.

\section{CME Mass \& Kinematics}
\label{sec:mass}
Some current CME catalogues provide estimates of CME mass and kinematics based on the plane-of-sky (POS) approximation. The estimate of the true 3D distribution of a CME in the extended inner corona enables a better approximation of mass and kinematics.

\subsection{CME mass}
For the standard routines provided in the LASCO (C3 only) and SECCHI COR Solarsoft routines, the brightness of the CME is estimated through a running base-difference (the subtraction of a pre-event image). This brightness is assumed to come solely from electrons embedded in the POS, which allows convenient inversion of calibrated brightness into a total electron number, easily converted to an estimate of mass assuming charge equality and a set abundance of helium. In the absence of calibration, and, more importantly, image processing errors, this mass estimate is a lower limit. If the CME (or most of the CME) is not embedded in the POS, the true mass must be higher. For a series of images containing a CME propagating through the FOV, several mass estimates may be made, and the maximum value found across the time series will give the best mass estimate. \citet{vourlidas2000} provides greater detail of the procedure and inherent uncertainties.

The 3D detection maps not only give a good estimate of the true CME direction of propagation, but also an estimate (albeit uncertain, particularly in longitude) of the spatial size of the CME. By assuming that the CME signal comes from only within the longitude-latitude areas of CME detection, the brightness of individual calibrated images from LASCO C3 are backprojected and inverted. A coarse voxel grid, filling the longitudinal and latitudinal areas of detected CME activity, and extending from heights 6.0 to 18\Rs, is used. Individual LASCO C3 images are calibrated according to the standard Solarsoft routines. The pixels in the images containing CMEs are backprojected through the CME voxel grid, and the brightness inverted to electron density using standard formulations \citep[e.g.][]{quemerais02}. Each voxel is large enough to contain backprojected values from several pixels, giving an initial mean and standard deviation. These are recorded, and the procedure repeated for all images comprising the set spanning a time window of several hours following the start time used for the original 3D detections. Density is converted to mass by multiplying by voxel volume, and a total mass estimate gained from summing. 

\begin{figure}[!t]
\centering
\includegraphics[width=8.5cm]{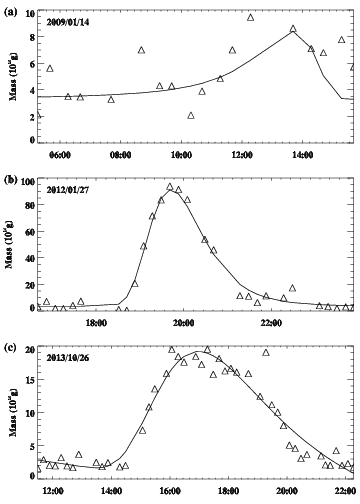}
\caption{Estimates of mass as a function of time for the CMEs of (a) 14/01/2009, (b) 27/01/2012, and (c) 26/10/2013. The symbols show each estimate for a time series of LASCO C3 observations, whilst the solid line shows a fitted skewed Gaussian plus linear background, as described in the text.}
\label{cme_mass}
\end{figure}

Based on current typical image cadences of LASCO C3, the procedure typically results in a set of at least 20 individual estimates of mass for different times. Mass estimates as a function of time are shown in figure \ref{cme_mass}. The mass estimate increases and decreases from a broad peak as the CME moves through the FOV. A typical approach in estimating CME mass is to take the maximum value as the best estimate. A better approach is to fit the broad peaks to a function of time, which allows an estimate of error. The function we choose is the sum of a skewed Gaussian and linear background. The amplitude of the Gaussian gives the CME mass, which means that the low background measurements in absence of the CME are absent from this estimate. Our final estimates of mass for the three case study CMEs presented in the previous section are $(5.1\pm1.7)\times10^{14}$g, $(8.5\pm0.4)\times10^{15}$g and $(2.0\pm0.2)\times10^{15}$g for 14/01/2009, 27/01/2012 and 26/10/2013, respectively. The CDAW estimates are $2.8\times10^{15}$g, $3.7\times10^{16}$g and $3.3\times10^{15}$g.

Considering the final combination of mass from a large number of voxels, each containing a large number of backprojected densities, a simplistic error propagation of the original standard deviations gives a very small error for the final masses ($\sim$1\%). However, this does not take into account systematic calibration errors and more importantly errors arising from the initial calculation of CME brightness based on image processing. Any estimation of the CME brightness involves separation of the CME signal from the background. The standard approach is a base-difference (subtracting a pre-CME image). This approach contains very large errors since (i) the background may change in response to the passage of large CMEs, and be included in the final mass estimate, (ii) the base image may itself contain previous CMEs leading to regions of negative mass in the difference image, and (iii) simple subtracting of images increases noise by a factor of approximately $\sqrt2$. The DST process used in this study helps avoid the errors associated with the second and third points, and is less sensitive to the first. A portion of the CME signal may be filtered out from the final CME images, leading to a low estimate. It is very difficult to quantify these errors since they differ from event to event depending on size, speed and the response of background quiescent structures to the passage of the CME. The mass estimate is therefore only an approximate indicator, yet gives a relative scale against which different CMEs may be compared. The same point is made succinctly by \citet{vourlidas2000}.


\subsection{CME kinematics}
For most studies and current catalogues, CME kinematics are also calculated using the POS approximation, where the position of the front edge of the CME is tracked in image (POS) coordinates as a function of time and fitted to a constant velocity or constant acceleration model. This measurement is less prone to inherent calibration and image processing errors than the mass estimate, but is also very much a lower limit on the true CME speed. The error becomes very large if the true CME direction of propagation is far from the POS, and becomes nonsensical to use in the case of halo or partial halo CMEs.

The CME POS kinematics are measured directly from the position of the front edge of the CME in both LASCO C3 and SECCHI COR2 A \& B. Given that what is taken as the front of the CME can often be the flank of a curved surface, the CME is assumed as a self-similarly expanding sphere. The geometry used is described by \citet{davies13}. The angular half width of the sphere is the width of region detected by ACT, which is an estimate for the spatial size of the CME. The sphere is assumed to expand with constant angular half width as it propagates also the centroid returned by ACT. Applying this geometry to the extracted kinematics yields the distance of the sphere front from the Sun. This is assumed as the true height of the CME at the given time. The height-time points are fitted to a linear model (constant speed) and second order model (constant acceleration). Errors are calculated for each parameter using the bootstrapping approach described by \citet{byrne2013}. This procedure gives up to three separate sets of kinematics for comparison against the values listed by CDAW. For the three case study CMEs, we find kinematic measurements from LASCO observations listed in Table 4, which also gives the CDAW POS estimates.

\begin{table}
\label{table:4}
\centering
\caption{ACT kinematics based on a self similarly expanding spherical geometry and bootstrapping using LASCO C2/C3 observations of the three case events. Also shown are the CDAW POS estimates for the same CMEs. v$_{a/c}$ velocity and a$_{a/c}$ acceleration where the subscript denotes ACT (a) or CDAW (c).}
\begin{tabular}{lrrr}
& 2009/01/14 & 2013/10/26 & 2012/01/27\\\hline
v$_{a}$ ($km.s^{-1}$) & 319$\pm$31 & 552$\pm$16 & 2395$\pm$28 \\
v$_{c}$ ($km.s^{-1}$) & 298 & 549 & 2508 \\\hline
a$_{a}$ ($m.s^{-2}$) & 11$\pm$0.7 & -3$\pm$0.2 & 20$\pm$1.7 \\
a$_{c}$ ($m.s^{-2}$) & -69.6 & -5.3 & 165.9 \\
\end{tabular}
\end{table}

The velocity for Event 1 (14/01/2009) was found to be higher than the value listed by CDAW. This is to be expected as POS approximations yield the lowest possible result. The velocity of Event 2 (26/10/2013) is very close to the CDAW result (within the given margin of error), which is again to be expected given how close this event propagates to the POS. CDAW lists Event 3 (27/01/2012) as propagating at 2508 km.s$^{-1}$. This velocity is extremely high, and is likely a case of POS estimates being unsuitable for halo eruptions. The velocity found by the present method is also high at 2395 km.s$^{-1}$. It is possible that this event is not well modelled by a spherical geometry.

\section{Summary}
\label{sec:summary}

In this paper, we present a new, automated method of detecting coronal mass ejections (CMEs) in three dimensions for the LASCO C2 and STEREO COR2 coronagraphs; the automated CME triangulation method. By triangulating isolated CME signal from the three coronagraphs over a sliding window of five hours, the most likely region through which CMEs pass at 5\Rs\ is identified. The technique presented is both quick and easily implemented, with no need for human input, and returns promising results with strong reliability so far. This research is therefore of great importance when attempting to avoid the potentially adverse effects of space weather.

For testing purposes, we apply the ACT method to a series of synthetic CMEs produced from a wireframe flux rope model, rendered from different phases of the STEREO mission. The angular difference between the input parameters of the flux rope model and the results of the ACT method range from 2.7$^{\circ}$ $\le$ $\sigma$ $\le$20$^{\circ}$. (Sections \ref{sec:testing} and \ref{sec:testing2}). When simulating the loss of one of the three observers, these values increase by up to 14.2$^{\circ}$. This represents a slight reduction in the accuracy of the technique, but not enough to render it obsolete should communications with STEREO-B never be restored. ACT is also demonstrated on three CMEs of different sizes and morphologies. The detection results are compared, where available, to those from a related technique (SSSE), extracted from HELCATS; or to the erupting coordinates on the solar surface of the triggering event (i.e. filament eruption), extracted from HEK. The results, of course, have a larger angular difference from the comparison data than the synthetic tests. A large part of the increase in $\sigma$ can be attributed to inherent uncertainties in the methods. Estimates of CME mass and kinematics are also calculated and are demonstrably reliable when compared with those listed by CDAW. An automated routine for detecting the CME leading front is in development and will be implemented into the kinematic analysis so that the entire procedure is fully automated. This body of work is not intended to be a rigorous comparison of various techniques, but introduces a simple, yet fully automated, method for detections of CMEs in 3D. A future study of a large survey of CMEs is planned to provide meaningful error estimates for ACT. Following this, ACT will be incorporated into the CORIMP database in the near future, enabling improved space weather diagnostics and forecasting.\\

\textbf{Acknowledgements:} Joe's work is conducted under an STFC studentship to the Solar Systems Physics group at Aberyswyth University. Huw is grateful for a research fellowship from the Leverhulme Foundation, which made this work possible. The authors also acknowledge our colleague Thomas Williams of Aberystwyth University, for inspiring the ``ACT'' name given to the method. The SOHO/LASCO data used here are produced by a consortium of the Naval Research Laboratory (USA), Max-Planck Insitut f{\"u}r Aeronomie (Germany), Laboratoire d'Astronomie (France), and the University of Birmingham (UK). SOHO is a project of international cooperation between ESA and NASA. The STEREO/SECCHI project is an international consortium of the Naval Research Laboratory (USA), Lockheed Martin Solar and Astrophysical Laboratory (USA), NASA Goddard Space Flight Center (USA), Rutherford Appleton Laboratory (UK), University of Birmingham (UK), Max-Planck Institut f{\"u}r Sonnen-systemforschung (Germany), Centre Spatial de Liege (Belgium), Institut d'Optique Th{\'e}orique et Appliqu{\'e}e (France), and Institut d'Astrophysique Spatiale (France). The SDO data used are provided courtesy of NASA/SDO and the AIA science team.

\bibliographystyle{plainnat}

Weblinks:

\textsuperscript{1} http://cdaw.gsfc.nasa.gov/CME$\_$list/

\textsuperscript{2} http://sidc.oma.be/cactus/

\textsuperscript{3} http://spaceweather.gmu.edu/seeds/

\textsuperscript{4} http://cesam.lam.fr/lascomission/ARTEMIS/

\textsuperscript{5} http://alshamess.ifa.hawaii.edu/CORIMP

\textsuperscript{6} http://www.helcats-fp7.eu/index.html

\textsuperscript{7} http://www.lmsal.com/hek/

\end{document}